\documentclass[english,preprint,tightenlines,eqsecnum,aps,prd,nofootinbib]{revtex4}
\usepackage[T1]{fontenc}
\usepackage[latin9]{inputenc}
\usepackage{amssymb}

\makeatletter
\@ifundefined{textcolor}{}
{%
 \definecolor{BLACK}{gray}{0}
 \definecolor{WHITE}{gray}{1}
 \definecolor{RED}{rgb}{1,0,0}
 \definecolor{GREEN}{rgb}{0,1,0}
 \definecolor{BLUE}{rgb}{0,0,1}
 \definecolor{CYAN}{cmyk}{1,0,0,0}
 \definecolor{MAGENTA}{cmyk}{0,1,0,0}
 \definecolor{YELLOW}{cmyk}{0,0,1,0}
}
\newenvironment{lyxlist}[1]
{\begin{list}{}
{\settowidth{\labelwidth}{#1}
 \setlength{\leftmargin}{\labelwidth}
 \addtolength{\leftmargin}{\labelsep}
 }}
{\end{list}}

\makeatother

\usepackage{babel}
\begin{document}

\title{Late-time decay of perturbations outside extremal charged black hole}

\author{Orr Sela}

\address{Department of physics, Technion-Israel Institute of Technology, Haifa
32000, Israel}
\begin{abstract}
We analyze the late-time decay of scalar perturbations in extremal
Reissner-Nordstrom spacetime. We consider individual spherical-harmonic
modes $l$ of a test massless scalar field, restricting our attention
to initial data of compact support, with generic regular behavior
across the horizon. We obtain a decay rate {\normalsize{}$\propto t^{-(2l+3)}$}
(just like in Schwarzschild) for incident waves scattered by the black
hole. However, for waves originating at the horizon's neighborhood
we obtain a slightly slower decay, {\normalsize{}$\propto t^{-(2l+2)}$.}
We discuss relations to previous works.
\end{abstract}
\maketitle

\section{Introduction}

When a black hole (BH) is perturbed by massless fields---either gravitational,
electromagnetic, or scalar---at late time the perturbations typically
decay as an inverse power of $t$ (the Schwarzschild time coordinate).
This phenomenon was found by Price \cite{Price,Price2} about four
decades ago in the case of a Schwarzschild BH. The goal of this paper
is to extend the large-$t$ analysis to an extremal Reissner-Nordstrom
(ERN) BH. 

For simplicity and concreteness, we shall focus throughout this paper
on the case of a massless, minimally coupled, test scalar field $\Phi$
with compact initial support and with generic regular behavior across
the horizon. In the Schwarzschild case, Price \cite{Price} found
a decay rate $t^{-2l-2}$ when an initial static moment is present
and a decay rate $t^{-2l-3}$ when the initial data have compact support,
where $l$ denotes the mode's multipolar number. Soon afterward, Bicak
\cite{Bicak} extended the analysis to the Reissner-Nordstrom (RN)
background. In the nonextremal case, he obtained (for his choice of
generic initial data) the same decay rate $t^{-2l-2}$ as in Schwarzschild
(for initial static moment). However, in the case of ERN Bicak found
a much slower decay rate $\propto t^{-l-2}$. 

To analyze the scattering problem, it is useful to introduce the tortoise
radial coordinate $r_{*}$ (defined below). The relevant domain then
extends from the horizon ($r_{*}\rightarrow-\infty$) to infinity
($r\simeq r_{*}\rightarrow\infty$). The scattering problem involves
an effective potential, a certain function $V_{l}(r_{*})$. Bicak
\cite{Bicak} observed that in the case of ERN, in both boundaries
$r_{*}\rightarrow\pm\infty$ this potential has the same asymptotic
behavior $V_{l}\simeq l(l+1)/r_{*}^{2}$ (this sharply contrasts with
the nonextremal case, wherein $V_{l}$ decays exponentially in $r_{*}$
on approaching the horizon). Couch and Torrence \cite{conformal}
later showed that this symmetry applies to all values of $r_{*}$(and
not only asymptotically), i.e. the function $V_{l}(r_{*})$ is symmetric
in $r_{*}$. Further discussions about this symmetry along with interesting
implications of it can be found in Refs. \cite{Bizon,Lucietti}. This
surprising mathematical symmetry connects the scattering dynamics
near the horizon to that at the weak-field region.

This symmetry of the potential was later employed by Blaksley and
Burko \cite{Burko}, who considered two special classes of initial
data: (i) compact initial support which does \emph{not} extend up
to the horizon, and (ii) initially-static multipoles that extend up
to the horizon (and up to future null infinity). Using the aforementioned
symmetry of $V_{l}(r_{*})$, Blaksley and Burko obtained decay rates
$t^{-2l-3}$ and $t^{-2l-2}$, respectively. Note that this in itself
does not conflict with Bicak's result (a decay $\propto t^{-l-2}$):
Bicak obtained his result for the case of \emph{generic} initial data,
which extend regularly across the horizon. On the other hand, the
special subclasses considered by Blaksley and Burko are nongeneric
and are both ``weaker'' than Bicak's case (in terms of the asymptotic
behavior of the initial data on approaching the horizon). 

In this paper we revisit the analysis of the ERN case, for the generic
class of initial data. Namely, we allow the initial data to approach
(and, in fact, to cross) the horizon in a generic regular manner.
We carry out the analysis entirely in the time domain. We use the
following strategy: First, we use the above mentioned symmetry of
$V_{l}(r_{*})$ to transform the strong-field near-horizon scattering
problem into a new, weak-field problem at large $r$. This allows
us to employ certain approximate methods previously developed for
analyzing wave dynamics in weak-field regions \cite{Price,Barack1}.
As a result, we get an approximate expression for $\phi$ (the inverted
field), valid for evaluation points in the domain of large $r$ and
for weak-field initial data. Then, transforming the resulting $\phi$
back into the original problem, we obtain an approximate expression
for the original field $\Phi$ valid for near-horizon evaluation points
and initial data. In a subsequent paper \cite{long} that I hope to
address, we shall employ an exact result obtained by Aretakis \cite{Aretakis},
concerning the behavior of certain derivatives of $\Phi$ along the
horizon, in order to verify the validity of our approximations and
to obtain the \emph{exact} (i.e. for generic compact initial data
which crosses the horizon regularly, not necessarily confined to the
near-horizon region) asymptotic expression for $\Phi$, although it
is still valid only for evaluation points in the vicinity of the horizon.
We will later use the ``late-time expansion'' method in order to
remove this restriction and get the exact large-$t$ asymptotic behavior
of $\Phi$ at \emph{any} fixed $r$ outside the BH. Even though the
details of the last two procedures are postponed to the next paper
\cite{long}, we will sketch very briefly the key ingredients and
main results of them in the last section of this paper.

Our final result is given in Eq. (\ref{eq:Final}): The field generically
decays at large $t$ as $t^{-2l-2}$, multiplied by a certain function
of $r$ (corresponding to a static solution).

We note that in the special cases considered by Blaksley and Burko
\cite{Burko}, their results are fully consistent with our analysis.
In particular, when an ingoing wave packet is scattered off the BH
(namely, no initial support at the horizon), the decay rate will be
$t^{-2l-3}$. Our analysis is also fully consistent with the results
obtained by Lucietti et al. \cite{Lucietti}, who numerically found
decay rates $\propto t^{-2l-2}$ for $l=0,1,2$.

\section{Field equation and initial-value setup}

\subsection{Background metric}

The ERN geometry is given in Schwarzschild coordinates by the line
element 
\[
ds^{2}=-(1-M/r)^{2}dt^{2}+(1-M/r)^{-2}dr^{2}+r^{2}d\Omega^{2},
\]
where $M$ is the BH mass, and $d\Omega^{2}\equiv d\theta^{2}+\sin^{2}\theta\, d\varphi^{2}$.
We focus here on the external domain, $r>M$. The tortoise coordinate
$r_{*}(r)$ is defined by $dr/dr_{*}=(1-M/r)^{2}$. Fixing the integration
constant by setting $r_{*}(2M)=0$ (for later convenience), we get
\begin{equation}
r_{*}=r-M-\frac{M^{2}}{r-M}+2M\ln(r/M-1).\label{eq:r*}
\end{equation}
This function diverges to $+\infty$ at $r\rightarrow\infty$ and
to $-\infty$ at $r\rightarrow M$, and vanishes at $r=2M$. We also
define the two null coordinates $(u,v)$ by setting $t=v+u$ and $r_{*}=v-u$,
namely:
\begin{equation}
v=(t+r_{*})/2,\: u=(t-r_{*})/2.\label{eq:uvDEFINED}
\end{equation}

\subsection{Field equation}

We consider a massless scalar field $\Phi$ satisfying the standard
wave equation 
\begin{equation}
\square\Phi\equiv\Phi_{\:\:;\alpha}^{;\alpha}=0.\label{eq:Basic_Field}
\end{equation}
We decompose $\Phi$ into spherical harmonics $Y_{lm}(\theta,\varphi)$
in the usual way, 
\[
\Phi=r^{-1}\sum_{lm}\psi_{l}(r,t)Y_{lm}(\theta,\varphi).
\]
The field equation (\ref{eq:Basic_Field}) then reduces to a certain
hyperbolic equation for each $l$: 
\begin{equation}
\psi_{l}''-\ddot{\psi}_{l}=V_{l}(r)\psi_{l},\label{eq:separated}
\end{equation}
where the prime and overdot, respectively, denote partial derivatives
with respect to $r_{*}$ and $t$, and the effective potential $V_{l}(r)$
takes the form %
\footnote{$r$ is to be regarded here as an implicit function of $r_{*}$.%
} 
\begin{equation}
V_{l}(r)=\left(1-\frac{M}{r}\right)^{2}\left[\frac{2M}{r^{3}}\left(1-\frac{M}{r}\right)+\frac{l(l+1)}{r^{2}}\right].\label{eq:potential}
\end{equation}

Our goal in the rest of the paper will be to analyze the asymptotic
behavior of the fields $\psi_{l}$ at large $t$, for appropriate
initial conditions.

\subsection{Characteristic initial-value problem}

We now set the characteristic initial-value problem, wherein the initial
value of $\psi_{l}$ is specified along two intersecting radial null
rays, $u=\mathrm{const}$ and $v=\mathrm{const}$. For convenience
we choose the vertex of these two rays to be at $r_{*}=0$ (but none
of our conclusions should depend on the vertex location). Without
loss of generality we also place the vertex at $t=0$, placing the
two initial null rays at $u=0$ and $v=0$. The initial data are thus
composed of the two initial functions $\psi_{v}(v)\equiv\psi_{l}(u=0,v)$
and $\psi_{u}(u)\equiv\psi_{l}(u,v=0)$. %
\footnote{We have omitted here the index $l$ from $\psi_{u}$ and $\psi_{v}$
for brevity, and we do so for a few other quantities defined below.%
} 

Note that $u\rightarrow\infty$ and $v\rightarrow\infty$ correspond
to the horizon and to future null infinity (FNI), respectively. Since
we only consider here wave dynamics outside the BH, we shall only
be concerned with the domain $0\leq u,v<\infty$. 

We shall only be interested in initial data of compact support; that
is, $\psi_{v}(v)$ vanishes beyond a certain value of $v$. Furthermore,
on physical grounds we shall restrict our attention to initial data
which are perfectly regular at the horizon (later we shall be more
specific about this aspect ). Each such set of initial data may be
expressed as a superposition of two components: %
\footnote{To this end, one may choose any two parameters $u_{1},u_{2}$ satisfying
$0<u_{1}<u_{2}$, and then choose any smooth (say $C^{\infty}$) ``transition
function'' $z(u)$ at $u\geq0$ which satisfies $z=1$ at $0\leq u\leq u_{1}$
and $z=0$ at $u\geq u_{2}$. Given any original initial-value set
($\psi_{u}(u),\psi_{v}(v)$), one then decomposes it into a type-I
component ($z(u)\psi_{u}(u),\psi_{v}(v)$) and a type-II component
($[1-z(u)]\psi_{u}(u),\psi_{v}=0$). These two components will indeed
fail to be analytic at $u=u_{1,2}$, but this will not pose any problem
because we only assume analyticity of initial conditions at the horizon. %
}
\begin{lyxlist}{00.00.0000}
\item [{(I)}] ``Off-horizon compact initial data''---namely, $\psi_{u}(u)$
(like $\psi_{v}$) is only supported at a certain finite range of
$u$ (implying that the initial support of $\psi_{l}$ is well separated
from the horizon). 
\item [{(II)}] ``Horizon-based initial data''---namely, $\psi_{v}(v)$
vanishes, and $\psi_{u}(u)$ is only supported in the range $u\geq w$,
for a certain $w>0$.
\end{lyxlist}
Note that a localized wave packet incident from infinity (and subsequently
scattered or absorbed by the BH) may be represented by initial data
of type I, whereas perturbations initiated at the horizon's neighborhood
(or during the collapse) are represented by type II. Owing to the
superposition principle, it will be sufficient to consider these two
scattering problems (type I and type II) separately. As discussion
below, initial data of type I yield a conventional late-time behavior
(as far as the large-$t$ decay at $r>M$ is concerned), qualitatively
the same as in the Schwarzschild case. On the other hand, type-II
initial data lead to a slower late-time decay, and hence will generically
dominate at late time---a phenomenon special to the ERN case. It is
therefore the second type---namely horizon-based initial data---that
will mostly concern us here (though later we shall also comment on
the generic situation, which is a superposition of I and II).

\subsection{Asymptotic behavior at the horizon}

As was already mentioned above, we consider here initial data which
are perfectly regular at the horizon. For simplicity, we shall actually
assume that the initial data are analytic across the horizon. This
means that for any monotonic parameter $\lambda(u)$ which regularly
parametrizes the null ray $v=0$, $\psi_{u}(\lambda)\equiv\psi_{u}(u(\lambda))$
will be analytic as $\lambda$ crosses the horizon (namely as $u\rightarrow\infty$).
A convenient choice of such a parameter is the area coordinate $r$,
which is a monotonically decreasing function of $u$ along any $v=\mathrm{const}$
ray {[}$dr/du=-(1-M/r)^{2}<0${]}. Furthermore, the perfect regularity
of the ``ingoing Eddington'' coordinate system ($v,r,\theta,\varphi$)
at the horizon indicates the regularity of $r$ as a parameter, on
crossing the horizon. 

Thus, to address the regularity of $\psi_{u}(u)$ at the horizon,
we express it as a function of the parameter $r$. Analyticity of
the initial function $\psi_{u}$ implies that this function admits
a Taylor expansion in $r$ in the neighborhood of $r=M$. We cast
this expansion in the form 
\begin{equation}
\psi_{u}(u)=c_{0}+c_{1}(r/M-1)+c_{2}(r/M-1)^{2}+...\label{eq:Horizon_Expansion}
\end{equation}
wherein, recall, $r$ is to be regarded as a function of $u$, evaluated
along the ray $v=0$ {[}namely $r=r(r_{*}=-u)$; Note that we have
replaced the original Taylor's expansion parameter $r-M$ by its dimensionless
counterpart $r/M-1$ for later convenience---this merely amounts to
absorbing a factor $M^{k}$ in the coefficient $c_{k}$.{]} 

Our goal in this paper is thus to analyze the late-time behavior of
the field $\psi_{l}(u,v)$ which evolves from regular ``horizon-based''
initial data---namely, $\psi_{v}(v)=0$, and $\psi_{u}(u)$ which
vanishes at $u<w$, and which on approaching the horizon admits the
regular expansion (\ref{eq:Horizon_Expansion}). To this end, we shall
next introduce a transformation which mathematically maps our near-horizon
strong-field problem to a new, more convenient problem in weak field.

\section{Inversion transformation}

The ERN spacetime admits a conformal transformation \cite{conformal}
which maps the horizon to infinity and vice versa. This \emph{inversion
transformation} maps $r_{*}$ to $-r_{*}$, which also corresponds
to the transformation 
\begin{equation}
r\rightarrow\frac{Mr}{r-M}\equiv T(r).\label{eq:mapping}
\end{equation}
The action of this inversion on the various independent variables
used below may be summarized by 
\begin{equation}
T(r_{*})=-r_{*}\:,\; T(t)=t\:,\; T(u)=v\:,\; T(v)=u.\label{eq:Tuvt}
\end{equation}

The crucial observation is that \emph{the effective potential is invariant
under this inversion}, namely $V_{l}(T(r))=V_{l}(r)$. Therefore,
given any solution $\psi_{l}(r,t)$ of the field equation (\ref{eq:separated}),
the inversion transformation produces a new solution $\hat{T}(\psi_{l}(r_{*},t))\equiv\psi_{l}(-r_{*},t)$.
In terms of other choices of the independent variables, the map $\hat{T}$
also takes the forms $\hat{T}(\psi_{l}(r,t))\equiv\psi_{l}(T(r),t)$,
and $\hat{T}(\psi_{l}(u,v))\equiv\psi_{l}(v,u)$ (namely, the arguments
$u,v$ are simply interchanged). Further discussions about this conformal
transformation along with interesting implications of it can be found
in Refs. \cite{Bizon,Lucietti}.

We use this inversion to map our original problem (horizon-based perturbations)
into a more familiar one, in which the perturbations are based in
the large-$r$ region. Thus, we define 
\[
\Psi_{l}(u,v)\equiv\hat{T}(\psi_{l}(u,v))=\psi_{l}(v,u).
\]
The inverted field $\Psi_{l}$ satisfies the same field equation (\ref{eq:separated},\ref{eq:potential})
as the original field $\psi_{l}$, namely
\[
\Psi_{l}''-\ddot{\Psi}_{l}=V_{l}(r)\Psi_{l}.
\]
The corresponding initial functions $\Psi_{v}(v)\equiv\Psi_{l}(u=0,v)$
and $\Psi_{u}(u)\equiv\Psi_{l}(u,v=0)$ are immediately obtained from
$\psi_{u}$,$\psi_{v}$ through $\Psi_{v}(v=p)=\psi_{u}(u=p)$ and
$\Psi_{u}(u=p)=\psi_{v}(v=p)$ (for any $p\geq0$).

\subsection*{Asymptotic behavior of inverted initial data}

As mentioned above, the $T$-inversion maps the horizon to FNI and
vice versa. The asymptotic behavior of the inverted initial function
$\Psi_{v}$ on approaching FNI ($v\rightarrow\infty$) will thus be
dictated by the horizon-regularity requirement (\ref{eq:Horizon_Expansion})
on $\psi_{u}$. The small near-horizon expansion parameter $r/M-1$
is $T$-mapped into the small near-FNI parameter
\begin{equation}
\varepsilon\equiv\frac{T(r)}{M}-1=\frac{M}{r-M}.\label{eq:epsilon}
\end{equation}
Thus, our inverted problem involves initial data 
\begin{equation}
\Psi_{u}(u)=0\label{eq:Psiu}
\end{equation}
and $\Psi_{v}(v)$ which vanishes at $v<w$, and which, on approaching
FNI, behaves as 
\[
\Psi_{v}=c_{0}+c_{1}\varepsilon+c_{2}\varepsilon^{2}+...
\]
(wherein $r$ is to be regarded as a function of $v$, evaluated along
$u=0$---namely, $r=r(r_{*}=v)$). However, since $\varepsilon=M/r+(M/r)^{2}+...$
is itself a regular function of $1/r$, $\Psi_{v}$ actually admits
a regular expansion at FNI in powers of $1/r$ :
\[
\Psi_{v}=\hat{c}_{0}+\hat{c}_{1}(M/r)+\hat{c}_{2}(M/r)^{2}+...
\]
where $\hat{c}_{j}$ are dimensionless coefficients which are obtained
from $c_{j}$ in a straightforward manner: for example, $\hat{c}_{0}=c_{0}$,
$\hat{c}_{1}=c_{1}$, $\hat{c}_{2}=(c_{2}+c_{1})$, etc. 

Now, it is convenient for later purposes to insert a global scale
parameter into the initial function $\Psi_{v}$. In order to do so,
we redefine the initial function in the following way: $\Psi_{v}(r)\rightarrow\Psi_{v}\left(\frac{M}{R}r\right)$
where here $\Psi_{v}$ is regarded as a function of $r$, and $R$
is the desired scale parameter. Moreover, we set the original $w$
to be equal to $M$, such that the $w$ of the new initial function
is equal to $R$. %
\footnote{Note, however, that even though now $w=R$, we may refer to this quantity
in different places along the paper as either $w$ or $R$. This is
mainly because in some places we want to emphasize that we use this
quantity to represent the edge of the support of the initial function
(along the initial ray), while in other places we want to emphasize
that we use it to represent the scale parameter. Note also that the
parameter $w_{1}$ (introduced below) is scaled too, similar to $R$. %
} The asymptotic expansion at FNI of the new initial function takes
the form 
\begin{equation}
\Psi_{v}=\hat{c}_{0}+\hat{c}_{1}(R/r)+\hat{c}_{2}(R/r)^{2}+...\;.\label{eq:r-powers}
\end{equation}
We deliberately chose the scale parameter to be a free parameter $R$
rather than just the mass $M$, because later we shall need the freedom
to scale these two parameters differently. In particular, the weak-field
limit---a crucial ingredient in the analysis below---will be conveniently
(and uniformly) achieved by setting $R\gg M$ (and correspondingly
$w\gg M$, as noted above). 

Owing to the superposition principle, it will of course be sufficient
to analyze the contribution emerging from the individual inverse powers
of $r$ in Eq. (\ref{eq:r-powers}). However, at this stage we still
keep the entire Taylor expansion (\ref{eq:r-powers}) as initial data.

Summarizing, we are considering the following initial-value problem
for the inverted field $\Psi_{l}$: $\Psi_{u}(u)$ entirely vanishes,
and $\Psi_{v}(v)$ is only supported at $v\geq w$ and behaves at
large $v$ as a regular Taylor expansion in $R/r$, as described in
Eq. (\ref{eq:r-powers}).

\section{Weak-field analysis}

We shall now focus our attention on the case $w\gg M$, %
\footnote{Note that $w\gg M$ is implemented by choosing the scale parameter
$R$ such that $R\gg M$.%
} wherein the initial data for the inverted field $\Psi_{l}$ are contained
in the domain $r\gg M$. This will allow us to employ weak-field methods,
which remarkably simplify the analysis. 

Note that in terms of the original field $\psi_{l}$, this assumption
amounts to assuming that the initial data along the incoming ray $v=0$
are confined to a narrow domain in the neighborhood of the horizon,
a domain characterized by $r-M\ll M$. However, throughout this and
the next two sections we are concerned with the inverted field $\Psi_{l}$---for
which our assumption $w\gg M$ amounts to weak-field initial data
(only later, in section 7, we transform our results back to the original
field $\psi_{l}$). 

In the analogous Schwarzschild problem, it was noticed long time ago
\cite{Price} that the inverse-power late time tails are dictated
by the weak-field dynamics, namely by scatterings at $r\gg M$. We
shall proceed now to extend this approach to ERN. To this end one
first need to obtain the large-$r$ asymptotic behavior of $r(r_{*})$
and the effective potential $V_{l}$. %
\footnote{Recall that in the field equation (\ref{eq:separated}) the effective
potential at the right-hand side is to be considered as a function
of $r_{*}$, through the implicit function $r(r_{*})$.%
} At $r\gg M$, the function $r(r_{*})$ takes the asymptotic form
\begin{equation}
r=r_{*}-2M\ln(r_{*}/M)+O(r_{*}^{\:0}).\label{eq:r(r*)}
\end{equation}
The effective potential is dominated at large $r$ by the centrifugal
term $l(l+1)/r^{2}$, and hence takes the asymptotic form
\begin{equation}
V_{l}(r_{*})=\frac{l(l+1)}{r_{*}^{\:2}}+4Ml(l+1)\frac{\ln(r_{*}/M)}{r_{*}^{\:3}}+O(r_{*}^{\:-3}).\label{eq:V_asymptotic}
\end{equation}
One can easily notice that this asymptotic expression, like Eq. (\ref{eq:r(r*)}),
is independent of the BH charge; hence, it is common to the Schwarzschild
and ERN cases. 

In the analysis of the Schwarzschild case, Price \cite{Price} realized
that the asymptotic form (\ref{eq:V_asymptotic}) of the potential
is sufficient to determine the late-time decay of the scalar perturbation,
regardless of the detailed form of $V_{l}(r)$ at smaller $r$ values.
We shall see below that this is also the case in ERN, though with
one important difference: Whereas in the Schwarzschild case the curvature-induced
term $\propto M\ln(r_{*}/M)r_{*}^{\:-3}$ is crucial for the tail
formation, in our case the tails are already formed at (and thus dominated
by) the leading-order, flat-space term $\propto l(l+1)r_{*}^{\:-2}$
(this results from the difference in the type of initial data along
$u=0$, namely compact versus noncompact support, as we discuss below.)
In this sense, the effectiveness of the weak-field approach is much
more dramatic in the ERN case.

\subsection{Iterative expansion}

The large-$r$ asymptotic form (\ref{eq:V_asymptotic}) motivates
an iterative analysis of the evolution of $\Psi_{l}$. This approach
was initiated by Price \cite{Price}, and proved to be extremely useful
in the Schwarzschild case. Specifically, we adopt here a slightly
modified variant of this iterative scheme, developed by Barack's \cite{Barack1},
and extend it to our situation of inverse-power initial data on FNI
approach. In this method one first decomposes $V_{l}(r_{*})$ into
the flat-space centrifugal potential
\begin{equation}
V_{0}(r_{*})=\frac{l(l+1)}{r_{*}^{\:2}}\label{eq:V0}
\end{equation}
and a residue 
\begin{equation}
\delta V(r_{*})\equiv V_{l}-V_{0}=4Ml(l+1)\frac{\ln(r_{*}/M)}{r_{*}^{\:3}}+O(r_{*}^{\:-3}).\label{eq:delta_V}
\end{equation}
We rewrite it as 
\begin{equation}
\delta V(r_{*})\cong4\,\eta\, V_{0}(r_{*})\label{eq:delta_V1}
\end{equation}
where 
\begin{equation}
\eta\equiv\frac{M}{r_{*}}\ln\frac{r_{*}}{M}.\label{eq:Eta}
\end{equation}
Note that $\eta$ is a small factor in weak field: it is of order---or
smaller than---
\begin{equation}
\eta_{w}\equiv(M/w)\ln(w/M)\ll1.\label{eq:Eta_w}
\end{equation}

Exploiting the smallness of $\delta V$, we now formally decompose
$\Psi_{l}$ as
\[
\Psi_{l}=\Psi_{(0)}+\Psi_{(1)}+\Psi_{(2)}+...
\]
where the terms $\Psi_{(n)}$ are defined as follows: $\Psi_{(0)}$
satisfies the flat-space $l$-mode wave equation 
\begin{equation}
\Psi_{(0)}''-\ddot{\Psi}_{(0)}-V_{0}(r_{*})\Psi_{(0)}=0\label{eq:Psi0_equation}
\end{equation}
(with standard flat-space regularity conditions at $r_{*}=0$). In
turn the fields $\Psi_{(n>0)}$ satisfy a hierarchy of similar wave
equations, but with a source term emerging from $\propto\delta V$:
\begin{equation}
\Psi_{(n)}''-\ddot{\Psi}_{(n)}-V_{0}(r_{*})\Psi_{(n)}=\delta V(r_{*})\Psi_{(n-1)}\quad\quad\quad(n>0).\label{eq:Psi_nEquation}
\end{equation}

At this point, a technical remark should be made about the above decomposition
of $V_{l}$ into $V_{0}$ and $\delta V$: We have treated here (and
similarly in few other places later in this section) the quantity
$\ln(r_{*}/M)$, which is indeed $>\ln(w/M)$, as a large number,
thereby neglecting terms $O(r_{*}^{\:-3})$ (with no logarithmic enhancement)
compared to those kept in $\delta V$. This setup is in principle
fine, and it is indeed sufficient for our purposes here (as we focus
later on $\Psi_{(0)}$ only). However, if one's goal is to obtain
the actual leading mass-induced corrections to the tail amplitudes
of $\Psi_{l}$ in ERN, this specific expansion scheme turns out to
be insufficient. Instead, one has to include all the \emph{nonlogarithmic}
$O(r_{*}^{\:-3})$ terms as well in $\delta V$ (namely, the right-hand
side of Eq. (\ref{eq:delta_V}) should take the form $[a\ln(r_{*}/M)+b]r_{*}^{\:-3}$,
with the appropriate coefficients $a,b$). The inclusion of these
nonlogarithmic $O(r_{*}^{\:-3})$ terms in $\delta V$ is necessary
for the following reason: When one calculates the $M$-induced corrections
to the tails at the anticipated leading order $(M/R)\ln(R/M)$, one
eventually finds that these corrections cancel out. Instead, the actual
leading-order $M$-induced corrections to the tails only appear at
nonlogarithmic order $O(M/R)$. However, in order to obtain the full
tail corrections at order $M/R$, one must include all $O(r_{*}^{\:-3})$
terms in $\delta V$, not only the logarithmic-enhanced ones. %
\footnote{It is interesting to note, however, that in the analogous problem
of Schwarzschild's late-time tails the status of those nonlogarithmic
terms is different: In that case, the dominant tails emerge from $\Psi_{(1)}$
(rather than $\Psi_{(0)}$), but the nonlogarithmic $O(r_{*}^{\:-3})$
terms in $\delta V$ do \emph{not} contribute at leading order.%
} 

Here, however, our goal is merely to obtain the leading-order late-time
tails---which (in the ERN case, and assuming $w\gg M$) emerge from
$\Psi_{(0)}$, at order $(M/R)^{0}$. The exposition of $\delta V$
was only made here for the conceptual presentation of the iterative
scheme, but we do not need to calculate $\Psi_{(1)}$---with or without
the nonlogarithmic terms.

\subsection{Initial data for the iteration fields}

We still need to specify the initial data for the various $\Psi_{(n)}$
fields. We choose here the simplest approach: We adopt $\Psi_{v}(v)$
as the initial function for $\Psi_{(0)}$ (along $u=0$), while setting
vanishing initial data for all other $\Psi_{(n>0)}$. 

At this point we must address a subtlety which has a profound effect
on the mathematical nature of the initial-value problem for the iteration
fields: The potential $V_{0}\propto r_{*}^{\:-2}$, designed for weak-field
analysis, is singular at $r_{*}=0$. This is to be contrasted with
the full effective potential $ $$V_{l}(r_{*})$, which is perfectly
regular at $r_{*}=0$ (and actually everywhere). Obviously the singularity
of $V_{0}$ leads to a similar singularity in $\delta V$, again at
$r_{*}=0$. This singularity is problematic, as it may potentially
cause undesired singularities in the iterative fields $\Psi_{(n)}$,
an issue which must be addressed. 

Consider first the evolution of $\Psi_{(0)}$. Here it turns out that
the singularity problem is naturally cured, yet it has an important
impact on the nature of the initial-value problem: For any initial
data along (the section $v\geq0$ of) the $u=0$ ray, which satisfy
appropriate regularity conditions as $v\rightarrow0$, there exists
a \emph{unique} regular (at $r_{*}=0$) solution of the field equation
(\ref{eq:Psi0_equation}) throughout the domain $v\geq u\geq0$ (namely,
the portion $r_{*}\geq0$ of $v\geq0$). As a consequence, the original
characteristic initial-value problem, which required initial data
(for $\Psi_{l}$) along two crossing null rays $u=0$ and $v=0$,
is actually replaced now by a mixed boundary/initial value problem:
One should only specify initial data for $\Psi_{(0)}$ along a \emph{single}
null ray $u=0$ (at $v\geq0$), and in addition demand regularity
along the timelike line $r_{*}=0$. This will uniquely determine $\Psi_{(0)}$
in the aforementioned triangle-like domain ($v\geq u\geq0$). Stated
in other words, the initial function $\Psi_{v}^{(0)}(v)\equiv\Psi_{(0)}(v;u=0)$
is sufficient, and no analogous function of $u$ is required as initial
data for $\Psi_{(0)}$. %
\footnote{Recall that this behavior of $\Psi_{(0)}$---namely, the existence
of unique regular solutions, and also the need for characteristic
initial data only on a single null ray---is a well-known feature of
the standard flat-space wave equation for a multipole $l$ of massless
scalar field. {[}To verify the connection, one should just view the
spatial variable ``$r_{*}$'' in Eq. (\ref{eq:Psi0_equation}) as
the standard radial coordinate ``$r$'' in a fiducial flat space.{]}%
}

The situation with $\Psi_{(n\geq1)}$ is more delicate. Since we approximate
$\delta V$ by its leading-order term $\propto r_{*}^{\:-3}\ln(r_{*}/M)$
(for the sake of simplifying the weak-field analysis), this expression
for $\delta V$ is even more singular than $V_{0}$ at $r_{*}\rightarrow0$.
As a consequence, the fields $\Psi_{(n\geq1)}$ might become singular
at $r_{*}=0$ for a sufficiently large $n$. Note, however, that this
divergence is merely an artifact resulting from our large-$r_{*}$
approximation for $\delta V$, Eq. (\ref{eq:delta_V}). This is a
useful approximation in the region mostly relevant to our analysis
(namely the weak-field domain), but it causes an undesired artifact
at small $r_{*}$. A natural way to avoid this problem is to simply
chop the expression (\ref{eq:delta_V}) for $\delta V$ at a certain
small $r_{*}$ value which we denote $r_{*}^{shell}$. This leads
to the ``shell model'' introduced and analyzed in Ref. \cite{Barack1}
(where $r_{*}^{shell}$ was taken to be of order $M$). This procedure
led to a well-defined expression (independent of $r_{*}^{shell}$)
for the late-time tail associated with $\Psi_{(1)}$ (and an analysis
regarding the smallness of the contributions coming from $\Psi_{(n>1)}$
was carried out).

As we already mentioned, the divergence of $V_{0}$ at $r_{*}=0$
changes the nature of the initial-value problem for $\Psi_{(0)}$.
The same applies to all iteration fields $\Psi_{(n)}$: We only need
initial conditions for $\Psi_{(n)}$ along the outgoing ray $u=0$
(which we simply take to be $\Psi_{(n>0)}=0$)---and require regularity
of $\Psi_{(n)}$ at $r_{*}=0$.

\subsection{The flat-space limit}

We are focusing here on the case $w\gg M$ (i.e. $M/R\ll1$). This
limit may be achieved by either increasing $w$ (by increasing $R$),
or by decreasing $M$ (fixing $M$ or $w$, respectively). Both procedures
are equivalent, they yield essentially the same evolving field $\Psi_{l}$.
In the following discussion we shall take the view-point that the
initial function $\Psi_{v}(v)$ is held fixed (in particular $w$
and $R$ are fixed), and we are considering the limit of decreasing
$M$.

In this limiting process, $\Psi_{(0)}$ remains fixed, as both its
field equation and its initial data are essentially independent of
$M$. %
\footnote{The dependence of $v(r)$ (along the ray $u=0$) on $M$ fades out
as $M\rightarrow0$.%
} 

In order to figure out the behavior of $\Psi_{(n>0)}$ in this limiting
process, we follow the lines of Ref. \cite{Barack1} (where the Schwarzschild
case was considered). First, we notice that the value of $\Psi_{(n)}$
is determined by the values of $\delta V\Psi_{(n-1)}$ taken in the
domain (i.e. the support) of the Green's function (constructed relatively
to the chosen evaluation point) of the operator of the left-hand side
of Eq. (\ref{eq:Psi0_equation}). The support of this Green's function
is not limited only to the weak-field domain, but also reaches to
the strong-field domain (and in particular reaches to $r_{*}=0$),
as shown and explained in detail in \cite{Barack1}. This means that
$\Psi_{(n)}$ will be influenced by sources located at any value of
$r_{*}$, and not only by large-$r_{*}$ sources. However, in \cite{Barack1},
Barack shows that (in Schwarzschild) the dominant (i.e. leading order
in $M/u\lesssim M/R\ll1$) contribution to $\Psi_{(n)}$ at FNI is
only due to sources located at large $r_{*}$ (i.e. $r_{*}\gg M$).
As a result, he shows that in the case $M/R\ll1$, (i) the iterative
series converges at null infinity at late time and (ii) the dominating
term of the expansion is the one of lowest possible $n$. Specifically,
Barack shows that $\Psi_{(n+1)}/\Psi_{(n)}\propto M/R$ (potentially
with logarithmic factors) at FNI. 

Similarly, in ERN (which has the same asymptotic structure as Schwarzschild
at $r_{*}\gg M$) we will regard the fields $\Psi_{(n)}$ evaluated
in the vicinity of FNI (late times and $r_{*}\gg M$) as originating
from sources located at large $r_{*}$. As a result, $\Psi_{(1)}$
decreases in the limit $M/R\ll1$, as we regard its source as coming
from the weak-field domain and as this source term is proportional
to $\delta V$, which in turn is $\propto\eta\lesssim\eta_{w}\ll1$,
cf. Eqs. (\ref{eq:Psi_nEquation},\ref{eq:delta_V1},\ref{eq:Eta_w}).
This argument applies to higher-order iterations as well (like in
Schwarzschild), suggesting that as $M$ decreases, $\Psi_{(n)}$ should
decrease as $(\eta_{w})^{n}$ (or roughly as $(M/R)^{n}$). We may
therefore naturally expect that at the limit $M\ll R$, the late-time
tails will be dominated by the lowest possible $n$---namely by the
flat-space field $\Psi_{(0)}$ (remember, though, that this approximation
for $\Psi_{l}$ is only valid for evaluation points in the vicinity
of FNI, i.e. at late times and in the domain $r_{*}\gg M$).

At this point it is instructive to compare our present situation to
the standard problem of late-time tails in Schwarzschild. In that
case too, for weak-field initial data, the late-time tails are dominated
by the minimal possible iteration order $n$. However, in the Schwarzschild
case, considering compact initial support (the case of greatest physical
motivation), it turns out that $\Psi_{(0)}$ yields no contribution
to the late-time field. This is directly related to the compact support
of the initial data. %
\footnote{In the other case of initial static moment, the initial data are not
compactly supported. Nevertheless, for the specific initial function
characterizing the initial static moment, it turns out that again
$\Psi_{(0)}$ has no contribution to the late-time field. This situation
changes when the initial function $\Psi_{v}$ is noncompactly supported
and decays as $v^{-k}$ (for a range of $k$ values), as we discuss. %
} In our problem, however---the inverted field $\Psi_{l}$ on the ERN
background---the initial data is noncompactly supported, resulting
in a nonvanishing tail contribution already for $n=0$, as we soon
show. Correspondingly, in our problem it will be $\Psi_{(0)}$ (rater
than $\Psi_{(1)}$) which dominates the late-time tails. For this
reason we shall devote the next two sections to study the evolution
of $\Psi_{(0)}$ and its late-time tail. 

It should be pointed out already at this stage that there is a potential
loophole in the above reasoning, concerning the dominance of $\Psi_{(0)}$
in the late-time tail: It is not difficult to imagine a situation
in which $\Psi_{(1)}$ is indeed suppressed globally as $M/w$ (or
$\eta_{w}$), and yet it may decay slower than $\Psi_{(0)}$ at large
$t$. However, there is evidence that this is not the case, which
we discuss in the Summary section. The first evidence is that $\Psi_{(1)}$
decays with the same inverse power of $t$ as $\Psi_{(0)}$, yet its
overall amplitude is suppressed by the small parameter $\sim M/w$.
The second and much stronger evidence emerges from the horizon's exact
conserved quantities (discussed in the Summary section) and suggests
that the mere effect of the higher-order iteration fields $\Psi_{(n>0)}$
is to modify the value of the constant prefactor in the resulting
expression for $\Psi_{(0)}$. We shall return to address this issue
in the Summary section.

Therefore, in the case $w\gg M$ considered here, it is indeed possible
to obtain the leading-order (in $\sim M/w$) of the late-time tails
from the flat-space field $\Psi_{(0)}$. This is the strategy that
we shall use in the rest of this paper.

\section{Analysis of the flat-space field $\Psi_{(0)}$}

We turn now to analyze the evolution of $\Psi_{(0)}$, governed by
Eq. (\ref{eq:Psi0_equation}). In several occasions it will be useful
to interpret this equation as that of a standard scalar field $\tilde{\Phi}$,
related to $\Psi_{(0)}$ through
\begin{equation}
\tilde{\Phi}\equiv\Psi_{(0)}Y_{lm}(\theta,\varphi)/r_{*},\label{eq:Phi_tild}
\end{equation}
which lives in a fiducial 3+1 flat background with standard Minkowski
spherical coordinates ($t,r_{*},\theta,\varphi$). (Here $r_{*}$
serves as the standard radial coordinate in this fiducial flat spacetime;
note that in Minkowski $r_{*}=r$.) The standard wave equation $\square\tilde{\Phi}=0$
then reduces to Eq. (\ref{eq:Psi0_equation}). 

For later convenience we also express the wave equation (\ref{eq:Psi0_equation})
in terms of the null coordinates $u,v$:
\begin{equation}
\Psi_{(0),uv}=-\frac{l(l+1)}{(v-u)^{2}}\Psi_{(0)}.\label{eq:Psi0,uv}
\end{equation}

\subsection{The general solution}

The general solution of this equation is known explicitly. It involves
two arbitrary functions $g(u)$ and $h(v)$, and it takes the form
\begin{equation}
\Psi_{0}(u,v)=\sum_{n=0}^{l}A_{n}^{l}\frac{g^{(n)}(u)+(-1)^{n}h^{(n)}(v)}{(v-u)^{l-n}},\label{eq:general_solution}
\end{equation}
where the upper index ``$(n)$'' denotes $n$th-order derivative
of the function with respect to its argument, and where $A_{n}^{l}$
are coefficients given by 
\begin{equation}
A_{n}^{l}=\frac{(2l-n)!}{n!(l-n)!}.\label{eq:Aln}
\end{equation}

If characteristic initial data for $\Psi_{0}$ were given in a local
``diamond'' which does not extend to $r_{*}=0$, the two functions
$g(u)$ and $h(v)$ could be chosen arbitrarily and independently.
However, when the domain of dependence contains the line $r_{*}=0$
(which is the case in our problem), for an arbitrary choice of $g(u)$
and $h(v)$ the solution (\ref{eq:general_solution}) generically
diverges at $r_{*}=0$, owing to the factor $(v-u)^{l-n}$ in the
denominator. %
\footnote{In the special case $l=0$, $\Psi_{0}$ itself is finite, yet $\tilde{\Phi}=\Psi_{(0)}/r_{*}$
generically diverges.%
} The most singular contribution, proportional to $(v-u)^{-l}$, emerges
solely from the $n=0$ term. In that term the numerator reads $g(u)+h(v)$.
The only way to avoid this divergence is to have the two functions
$g(u)$ and $h(v)$ cancel each other along the line $v=u$. This
cancellation occurs if and only if $g(u)$ and $h(v)$ have the same
functional form, that is, $g(x)=-h(x)\equiv f(x)$. Thus, the general
solution regular at $r_{*}=0$ takes the form 
\begin{equation}
\Psi_{0}(u,v)=\sum_{n=0}^{l}A_{n}^{l}\frac{f^{(n)}(u)-(-1)^{n}f^{(n)}(v)}{(v-u)^{l-n}}.\label{eq:general_regular}
\end{equation}

We shall refer to the function $f(x)$, which underlies the expression
in the right-hand side (through substitutions $x\rightarrow u$ and
$x\rightarrow v$), as the \emph{generating function}. Also, we shall
denote the differential operator at the right-hand side, acting on
the function $f$, by $\Delta(f)$.

Since the form of the general solution for $\Psi_{0}(u,v)$ is known,
the remaining challenge is to find the function $f$ which corresponds
to the specified initial function $\Psi_{v}(v)$. Namely, we seek
the function $f$ satisfying 
\begin{equation}
\Delta(f)(u=0,v)=\Psi_{v}(v).\label{eq:Delta}
\end{equation}
In the following subsections we shall proceed to construct the solution(s)
of this equation. To this end, we shall first discuss several properties
of the generating function $f$ and of the differential operator $\Delta$.

\subsection{General properties of the generating function $f$ }

We would first like to explore the space of homogeneous solutions
of Eq. (\ref{eq:Delta}), namely the functions $f$ satisfying $\Delta(f)(u=0,v)=0$.
Note that $\Psi_{v}(v)=0$ implies the vanishing of $\Psi_{0}(u,v)$
due to uniqueness (and vice versa---the vanishing of $\Psi_{0}(u,v)$
trivially implies $\Psi_{v}(v)=0$); hence, these are the functions
$f$ generating a trivial field $\Psi_{0}(u,v)=0$. 

We shall therefore focus on the projection of Eq. (\ref{eq:general_regular})
to $u=0$: 
\begin{equation}
\Psi_{v}(v)=\sum_{n=0}^{l}A_{n}^{l}\frac{f^{(n)}(0)-(-1)^{n}f^{(n)}(v)}{v^{l-n}}\equiv\Delta_{0}(f(v)),\label{eq:Delta0}
\end{equation}
which defines the differential operator $\Delta_{0}$ acting on functions
$f(v)$. Here $f^{(n)}(0)$ means $f^{(n)}(v)$ evaluated at $v=0$. 

Yet another relevant differential operator is the restriction of $\Delta_{0}$
to its local piece: 
\begin{equation}
D_{0}(f(v))\equiv\sum_{n=0}^{l}(-1)^{n+1}A_{n}^{l}\, v^{n-l}\, f^{(n)}(v),\label{eq:D_0}
\end{equation}
such that 
\begin{equation}
\Psi_{v}(v)=\Delta_{0}(f(v))=D_{0}(f(v))+\sum_{n=0}^{l}A_{n}^{l}\, v^{n-l}\, f^{(n)}(v=0).\label{eq:D0(f)}
\end{equation}
Recall the difference between these three linear operators: $\Delta$
is a \emph{partial} differential operator, $\Delta_{0}$ is an \emph{ordinary}
differential operator, but yet a nonlocal one (it involves the values
of $f^{(n)}$ at $v=0$), but $D_{0}$ is a more standard, local,
ordinary differential operator. 

The space of homogeneous solutions of these differential operators
was analyzed in Ref. \cite{Barack1}. Here we merely mention the results: 
\begin{lyxlist}{00.00.0000}
\item [{(I)}] The basis of homogeneous solutions of $D_{0}$ is 
\begin{equation}
v^{j}\;,\qquad l+1\leq j\leq2l.\label{eq:Homog_D0}
\end{equation}

\item [{(II)}] The basis of homogeneous solutions of $\Delta_{0}$ (or,
equivalently, $\Delta$) is larger:
\begin{equation}
v^{j}\;,\qquad0\leq j\leq2l.\label{eq:Homog_Delta0}
\end{equation}

\end{lyxlist}
Recall that $D_{0}$ is a rather standard (i.e. local) $l$th-order
differential operator; hence, it must admit $l$-independent homogeneous
solutions, as in Eq. (\ref{eq:Homog_D0}). It is a bit surprising
that the set (\ref{eq:Homog_Delta0}) is larger, including $2l+1$
independent functions, even though $\Delta_{0}$ too is an $l$th-order
differential operator. This abnormally-large basis is only possible
due to the nonlocal nature of the operator $\Delta_{0}$. 

This somewhat confusing structure of the space of homogeneous solutions
associated with $\Delta_{0}$ and $D_{0}$ is most easily illustrated
in the $l=0$ case. In that case, $\Psi_{0}(u,v)=f(u)-f(v)\equiv\Delta(f)$,
then $\Psi_{v}(v)=f(v=0)-f(v)\equiv\Delta_{0}(f)$, and $D_{0}(f)=-f(v)$
is a trivial (i.e. zero-order) differential operator. Obviously $D_{0}$
has no (nonvanishing) homogeneous solutions, yet $\Delta_{0}$ does
have one nonvanishing homogeneous solution: $f(v)=\mathrm{const}.$
For this choice, $\Psi_{0}(u,v)=f(u)-f(v)$ indeed vanishes.

Note also that every homogeneous solution of $D_{0}$ must also be
a homogeneous solution of $\Delta_{0}$, as can be seen by comparing
the two bases (\ref{eq:Homog_D0},\ref{eq:Homog_Delta0}). This must
indeed be the case, by virtue of Eq. (\ref{eq:D0(f)}), because for
any member of the smaller basis (\ref{eq:Homog_D0}), all functions
$f^{(n)}(v)$ with $0\leq n\leq l$ vanish at $v=0$.

\subsubsection{Representative generating function}

For a given initial function $\Psi_{v}(v)$, the general solution
of the inhomogeneous equation $\Delta_{0}(f(v))=\Psi_{v}(v)$, for
$f(v)$, is any specific inhomogeneous solution, plus any homogeneous
solution. This is an infinite set, a $2l+1$-parameter family of solutions.
We shall prescribe here the construction of a \emph{single} solution
$f(v)$, for a given function $\Psi_{v}(v)$---which we shall call
the \emph{representative solution}. 

We shall assume that $\Psi_{v}(v)$ vanishes as $v\rightarrow0$.
This is trivially satisfied by the initial functions considered here,
because they were chosen to vanish at $v<w$. We note, however, that
even without this choice of restricted support, $\Psi_{v}(v\rightarrow0)$
should vanish in order for $\tilde{\Phi}\propto\Psi_{(0)}/r_{*}$
to be bounded at $u=v=0$. 

We define our representative solution $f(v)$ as follows: It is the
solution of the $l$th order ODE, 
\begin{equation}
D_{0}(f(v))=\Psi_{v}(v),\label{eq:f_ODE}
\end{equation}
subject to the initial conditions 
\begin{equation}
f^{(n)}(v=0)=0\;,\qquad0\leq n\leq l-1.\label{eq:Init_cond}
\end{equation}
Notice that by virtue of these two equations, $f^{(l)}$ vanishes
at $v=0$. Therefore, each of the terms in the sum in the right-hand
side of Eq. (\ref{eq:D0(f)}) vanishes, yielding the required relation
$\Delta_{0}(f(v))=\Psi_{v}(v)$. We shall denote this representative
generating function by $f_{r}(v)$. 

For the class of initial functions $\Psi_{v}(v)$ considered here,
which vanish at $v<w$, $f_{r}(v)$ too will vanish at $v<w$. Therefore,
the representative solution will take the form
\begin{equation}
f_{r}(v)=\Theta(v-w)f_{w}(v),\label{eq:f_w}
\end{equation}
where $\Theta$ denotes the standard step function, and $f_{w}(v)$
is defined to be the solution of the ODE (\ref{eq:f_ODE}) with the
appropriate initial conditions at $v=w$: 
\begin{equation}
f^{(n)}(v=w)=0\;,\qquad0\leq n\leq l-1.\label{eq:Init_cond_w}
\end{equation}

\subsubsection{Compactly supported initial data}

Although in our (inverted) ERN case the function $\Psi_{v}(v)$ is
of noncompact support, it will be useful to consider here the other
case, in which the initial function $\Psi_{v}(v)$ is of compact support.
Let us assume thus that the support of $\Psi_{v}(v)$ is restricted
to the range%
\footnote{The restriction $v>w$ can be relaxed if we assume the regularity
condition $\Psi_{v}(v\rightarrow0)=0$.%
} 
\begin{equation}
w<v<v_{max}.\label{eq:Compact_Support_range}
\end{equation}

Consider now the expression (\ref{eq:f_w}) for $f_{r}(v)$. Since
$\Psi_{v}$ vanishes at $v>v_{max}$, throughout this domain $f_{r}(v)=f_{w}(v)$
will be a certain homogeneous solution, which we denote $f_{hom}(v)$;
namely, it satisfies $D_{0}(f_{hom}(v))=0$. We can therefore express
$f_{r}(v)$ as 
\begin{equation}
f_{r}(v)=\Theta(v_{max}-v)\left[\Theta(v-w)f_{w}(v)\right]+\Theta(v-v_{max})f_{hom}(v).\label{eq:f_r_compact}
\end{equation}
By virtue of Eq. (\ref{eq:Homog_D0}), $f_{hom}$ must be a superposition
of the form 
\begin{equation}
f_{hom}(v)=\sum_{j=l+1}^{2l}a_{j}\, v^{j}\label{eq:f_hom}
\end{equation}
with certain coefficients $a_{j}$ (whose values will not concern
us). From Eq. (\ref{eq:Homog_Delta0}), such a superposition also
satisfies $\Delta(f_{hom})(u,v)=0$.

Consider now the behavior of $\Psi_{0}$ in the portion $u>v_{max}$
of $r_{*}>0$. Any point in this domain satisfies $v_{max}<u<v$.
Therefore, in the right-hand side of Eq. (\ref{eq:general_regular}),
the functions $f^{(n)}(u)$ and $f^{(n)}(v)$ may all be replaced,
respectively, by $ $$f_{hom}^{(n)}(u)$ and $f_{hom}^{(n)}(v)$,
implying that 
\begin{equation}
\Psi_{0}(u,v)=\Delta(f_{hom})(u,v)=0\qquad\qquad(v>u>v_{max}).\label{eq:vanishing}
\end{equation}
We conclude that if the initial function $\Psi_{v}(v)$ is of compact
support, restricted to $v\leq v_{max}$, then $\Psi_{0}$ strictly
vanishes throughout the domain $u>v_{max}$ (of $r_{*}>0$). In particular,
in this case $\Psi_{0}$ will have no contribution whatsoever to the
late-time tails---either at Future null infinity or along $r=\mathrm{const}$. 

Note that this is precisely the reason why in the usual Schwarzschild
problem, with compactly-supported initial data, no late-time contribution
emerges from $\Psi_{0}$, hence the dominant contribution to the late-time
tails emerges from $\Psi_{(1)}$ (however, in our inverted ERN problem,
the situation is different due to the noncompactness of the initial
support).

\section{Construction of $\Psi_{0}$$ $ for inverse-power initial data }

\subsection{Simplifying the initial-value setup}

The initial conditions for $\Psi_{0}$, the function $\Psi_{v}(v)$,
was given in Eq. (\ref{eq:r-powers}). We shall now proceed to simplify
our initial-value setup, as we now describe. 

Recall that we have assumed analyticity of $\psi_{u}(r)$ across the
horizon, which in turn implies that the inverted field $\Psi_{v}$
is analytic in $1/r$ for a sufficiently large $r$. In particular,
the series (\ref{eq:r-powers}) converges at a sufficiently large
$r$. However, we have also assumed that $\Psi_{v}(v)$ vanishes at
$v\leq w$ (for some $w>0$). In fact this implies that $\Psi_{v}(v)$
is not everywhere analytic (neither in $v$ nor in $r$). But nevertheless
we assume that it is everywhere $C^{(\infty)}$. 

Thus, our initial function $\Psi_{v}(v)$ presumably admits the following
properties: (i) it is $C^{(\infty)}$ everywhere (in $r$ and, hence,
in $v$ too), (ii) it vanishes at $v\leq w$, and (iii) throughout
$v\geq w_{1}$ (for a certain $w_{1}\geq w$), it is given by the
convergent series (\ref{eq:r-powers}). 

To simplify the initial-value setup, we first note that owing to the
superposition principle, it will be sufficient to consider the contributions
emerging from individual terms $\hat{c}_{k}(R/r)^{k}$ separately.
Thus, we shall be concerned with an initial function which at $v\geq w_{1}$
takes the form

\begin{equation}
\Psi_{v}^{k}=\hat{c}_{k}(R/r)^{k}.\label{eq:single-power}
\end{equation}
In addition, $\Psi_{v}^{k}$ is assumed to vanish at $v\leq w$; this
implies that in the domain $w\leq v\leq w_{1}$, $\Psi_{v}^{k}$ will
inevitably deviate from Eq. (\ref{eq:single-power}); i.e., it is
not analytic. We shall assume that $\Psi_{v}^{k}$ is $C^{(\infty)}$.
\footnote{It is not difficult to construct such a $C^{(\infty)}$ extension
of (\ref{eq:single-power}) to the domain $ $$v\leq w_{1}$, as follows:
First, choose a parameter $w\leq\tilde{w}<w_{1}$ such that the sum
(\ref{eq:r-powers}) still converges to $\Psi_{v}$ at $w=\tilde{w}$.
Let $p(r)$ be any smooth transition function, that is, a monotonically
increasing $C^{(\infty)}$ function satisfying $p=0$ at $v\leq\tilde{w}$
and $p=1$ at $v\geq w_{1}$. Let us define $\tilde{\Psi}_{v}=\sum_{k}\hat{c}_{k}(R\, p/r)^{k}$
and $\Delta\Psi_{v}=\Psi_{v}-\tilde{\Psi}_{v}$. Notice that $\Delta\Psi_{v}$
is $C^{(\infty)}$ and is only supported at $w\leq v\leq w_{1}$.
We now define $\Psi_{v}^{k}=\hat{c}_{k}(R\, p/r)^{k}+\tilde{c}_{k}\Delta\Psi_{v}$,
where $\tilde{c}_{k}=\hat{c}_{k}(R/r_{1})^{k}/\Psi_{v}(v=w_{1})$
and $r_{1}\equiv r(v=w_{1})=r(r_{*}=w_{1})$. Note that $\sum_{k}\tilde{c}_{k}=1$.
It is not difficult to show that (i) $\Psi_{v}^{k}$ vanishes at $v\leq w$,
(ii) it coincides with Eq. (\ref{eq:single-power}) at $v\geq w_{1}$;
(iii) it is $C^{(\infty)}$ everywhere, and (iv) $\sum_{k}\Psi_{v}^{k}=\Psi_{v}$. %
} 

Next, to further simplify the analysis, we shall approximate $r$
in Eq. (\ref{eq:single-power}) by $r_{*}$. This simplification is
justified by our weak-field approximation. Recall that the relative
difference between $r$ and $r_{*}$ is $\propto M/r$ (setting aside
logarithmic corrections), which is negligible in the relevant domain
($r\gtrsim R\gg M$). 

We are thus led to consider the initial functions $\hat{\Psi}_{v}^{k}(v)$
(for all integers $k\geq0$), which are presumably $C^{(\infty)}$,
supported at $v\geq w$ only, and which satisfy $\hat{\Psi}_{v}^{k}=\hat{c}_{k}(R/r_{*})^{k}$
at $v\geq w_{1}$, namely

\begin{equation}
\hat{\Psi}_{v}^{k}(v)=\hat{c}_{k}(R/v)^{k}\label{eq:single-power-r*}
\end{equation}
at $v\geq w_{1}$. %
\footnote{The difference between $\hat{\Psi}_{v}^{k}\propto(R/r_{*})^{k}$ and
the ``true'' initial data $\Psi_{v}^{k}\propto(R/r)^{k}$, being
proportional to $M/R$ (aside from logarithmic corrections), should
naturally be considered as a seed for $\Psi_{(1)}$. Here, however,
we are only concerned with $\Psi_{(0)}$; hence, this difference between
$\hat{\Psi}_{v}^{k}$ and $\Psi_{v}^{k}$ will not concern us.%
}

\subsection{Generating function for inverse-power initial data}

In order to analyze the $\Psi_{(0)}$ field evolving from the above
initial function $\hat{\Psi}_{v}^{k}(v)$, we first construct the
corresponding representative generating function $f_{r}(v)$, which
satisfies the inhomogeneous ODE $D_{0}(f_{r}(v))=\hat{\Psi}_{v}^{k}(v)$
with the initial conditions (\ref{eq:Init_cond}). Since the space
of homogeneous solutions is known, Eq. (\ref{eq:Homog_D0}), all we
need at this stage is to construct a single inhomogeneous solution,
which we denote $f_{inh}(v)$. As we explain in the next subsection,
the late-time behavior of the field $\Psi_{(0)}$ depends only on
the values that $f_{r}(v)$ takes in the range $v\geq w_{1}$. Therefore,
we shall focus on finding $f_{inh}(v)$ {[}and $f_{r}(v)${]} in that
range (where $\hat{\Psi}_{v}^{k}(v)=\hat{c}_{k}(R/v)^{k}$). The structure
(\ref{eq:D_0}) of the operator $D_{0}$ immediately suggests the
existence of a specific solution of the form $f_{inh}(v)=\mathrm{const}\cdot v^{l-k}$
(in $v\geq w_{1}$). A direct substitution yields 
\begin{equation}
f_{inh}(v)=\left(\alpha\hat{c}_{k}R^{k}\right)v^{l-k}\qquad\quad(v\geq w_{1})\label{eq:f_inh}
\end{equation}
where $\alpha\equiv\alpha(l,k)$ is defined by
\[
1/\alpha=\sum_{n=0}^{l}(-1)^{n+1}A_{n}^{l}\prod_{j=0}^{n-1}(l-k-j)=-\sum_{n=0}^{l}A_{n}^{l}\prod_{j=0}^{n-1}(j+k-l).
\]
Using MATHEMATICA, one can easily find the more explicit expression
\begin{equation}
\alpha=-\frac{k!}{(k+l)!}.\label{eq:alpha}
\end{equation}

Once the specific inhomogeneous solution is known, the general solution
of the ODE (\ref{eq:f_ODE}) is obtained by adding a general homogeneous
solution of the form (\ref{eq:f_hom}). Therefore, the representative
generating function (in the range $v\geq w_{1}$) takes the form
\begin{equation}
f_{r}(v)=\left(\alpha\hat{c}_{k}R^{k}\right)v^{l-k}+\sum_{j=1}^{l}c_{j}v^{l+j}\qquad\quad(v\geq w_{1}),\label{eq:f_r_powers}
\end{equation}
with certain coefficients $c_{j}$. These coefficients are in principle
determined by integrating the ODE (\ref{eq:f_ODE}) with the initial
conditions (\ref{eq:Init_cond}), hence they depend on the specific
form of $\hat{\Psi}_{v}^{k}(v)$ at $v<w_{1}$. However, for the analysis
below we shall not need these coefficients.

\subsection{Late-time behavior of \textmd{\normalsize{}$\Psi_{0}$ }}

\subsubsection{General expression}

Consider now the behavior of $\Psi_{0}$ in the range $v>u\geq w_{1}$.
From the form (\ref{eq:general_regular}) of the general solution,
it is obvious that only the behavior of the generating function at
$v\geq w_{1}$ is relevant. Since the homogeneous piece $\sum c_{j}v^{l+j}$
yields no contribution to $\Psi_{0}$ {[}cf. Eq. (\ref{eq:Homog_Delta0}){]},
the late-time behavior of $\Psi_{0}$ is (precisely)
\begin{equation}
\Psi_{0}(u,v)=\left(\alpha\hat{c}_{k}R^{k}\right)\Delta\left(x^{l-k}\right)\qquad\quad(v>u\geq w_{1}),\label{eq:general_late}
\end{equation}
where, recall,
\begin{equation}
\Delta\left(x^{l-k}\right)\equiv\sum_{n=0}^{l}A_{n}^{l}\left[\frac{d^{n}}{du^{n}}(u^{l-k})-(-1)^{n}\frac{d^{n}}{dv^{n}}(v^{l-k})\right](v-u)^{n-l}.\label{eq:late_explicit}
\end{equation}

As was discussed in the previous section, $\Delta\left(x^{j}\right)$
vanishes for all $0\leq j\leq2l$, cf. Eq. (\ref{eq:Homog_Delta0}).
It immediately follows that at late time (namely, $ $$u\geq w_{1}$),
$\Psi_{0}$ vanishes for any $k\leq l$. We only need to consider
the contributions from $k>l$. 

This last result---the strict vanishing of late-time $\Psi_{0}(u,v)$
for all $k\leq l$---involves a subtlety: If $\Psi_{0}$ vanishes,
why is $\hat{\Psi}_{v}^{k}(v)$ nonvanishing? To address this question
we must distinguish between two different situations: (i) In the situation
described above, the initial function $\hat{\Psi}_{v}^{k}(v)$ is
$\propto v^{-k}$ only at $v\geq w_{1}$ (for some $w_{1}>0$). In
that case, $\Psi_{0}(u,v)$ vanishes at $v>u\geq w_{1}$, but \emph{not}
necessarily in the domain $w_{1}>u>0$. This allows continuity of
$\Psi_{0}$ on approaching $u=0$. (ii) The more subtle case is the
one in which $w_{1}\rightarrow0$, namely, $\hat{\Psi}_{v}^{k}\propto v^{-k}$
on the entire (portion $v>0$ of the) ray $u=0$. In that case, the
above result would mean that $\Psi_{0}(u,v)=0$ on the entire (portion
$r_{*}>0$ of the) domain $u\geq0$---which would conflict with the
nonvanishing value of $\hat{\Psi}_{v}^{k}$ at $u=0$. The resolution
of this conflict is a bit tricky: In this situation of $\hat{\Psi}_{v}^{k}\propto v^{-k}$
extending all the way to $v=0$, $\hat{\Psi}_{v}^{k}=\Psi_{0}(u=0)$
diverges at the origin ($u=v=0$). This divergence violates the regularity
condition that we have imposed along the central worldline $r_{*}=0$.
Thus, in this case there is a conflict between (I) the form of the
initial function at $v\rightarrow0$ , (II) the assumed regularity
condition at $r_{*}=0$, and (III) the assumption of continuity of
$\Psi_{0}(u,v)$ on approaching $u\rightarrow0$. The above construction
of $\Psi_{0}(u,v)$ was based on Eq. (\ref{eq:general_regular}),
which in turn was based on the assumption of regularity at $r_{*}=0$,
hence continuity at $u\rightarrow0$ has been sacrificed.

\subsubsection{Decay rate along future null infinity}

Consider next the behavior of $\Psi_{0}$ at the limit $v\rightarrow\infty$
(at fixed $u\geq w_{1}$), which characterizes the approach to future
null infinity (FNI). In the $v$-derivatives term in the squared brackets
in Eq. (\ref{eq:late_explicit}), for each $n$ the dominant contribution
(in terms of expansion in $1/v$) is $\propto v^{-k}$ and will hence
vanish at FNI (recall that the relevant values of $k$ are $k>l$).
In the $u$-derivatives term, the dominant contribution of the $n$th
term is $\propto v^{n-l}$, which vanishes for all $n<l$. The only
nonvanishing contribution to $\Delta\left(x^{l-k}\right)$ at FNI
is thus the $n=l$ term, which---since $A_{l}^{l}=1$--- is just $(d/du)^{l}(u^{l-k})$.
It immediately follows that $\Delta\left(x^{l-k}\right)$ vanishes
for any $0<k\leq l$. Calculation yields
\begin{equation}
\Delta\left(x^{l-k}\right)=\frac{d^{l}}{du^{l}}(u^{l-k})=p_{k}u^{-k}\qquad\quad(v\rightarrow\infty,\; u\geq w_{1}),\label{eq:Delta(x^l)}
\end{equation}
where
\begin{equation}
p_{k}=(-1)^{l}\,\prod_{j=0}^{l-1}(j+k-l)=(-1)^{l}\,\prod_{j=1}^{l}(k-j).\label{eq:Ak_1}
\end{equation}
The last expression shows at once that $p_{k}$ vanishes for any $0<k\leq l$.
This again demonstrates that---at least at FNI---$\Psi_{0}$ vanishes
throughout $u\geq w_{1}$ for any $0<k\leq l$. It confirms the more
general observation made above, that the late-time contributions to
$\Psi_{0}$ emerge only from $k>l$ terms. %
\footnote{The $k=0$ case requires a special treatment. (Nevertheless, the vanishing
of the $k=0$ contribution at $u\geq w_{1}$ follows from the more
general observation which we have just mentioned.) %
} Noting this range of $k$ values, it will be more convenient to express
$p_{k}$ as 
\begin{equation}
p_{k}=(-1)^{l}\frac{(k-1)!}{(k-l-1)!}\qquad\quad(k>l).\label{eq:Ak_2}
\end{equation}
Substituting back in Eqs. (\ref{eq:Delta(x^l)}) and (\ref{eq:general_late})
we find (for $k>l$ only) 
\begin{equation}
\Psi_{0}(u,v)=\left(\beta\hat{c}_{k}R^{k}\right)u^{-k}\qquad\quad(v\rightarrow\infty,\; u\geq w_{1}),\label{eq:general_late-1}
\end{equation}
where $\beta\equiv\beta(l,k)$ is given by
\[
\beta=(-1)^{l+1}\frac{(k-1)!}{(k-l-1)!}\frac{k!}{(k+l)!}=(-1)^{l+1}\frac{k\left[(k-1)!\right]^{2}}{(k-l-1)!(k+l)!}.
\]

While the specific value of the coefficient $\beta$ is not so crucial,
the important observation is that for any $k>l$ this coefficient
is nonvanishing. Recalling the vanishing of all $k\leq l$ contributions,
we conclude that the late-time field at FNI is dominated by the $k=l+1$
tail, and it hence decays as $u^{-l-1}$. 

To summarize this subsection, we want to write the dominant late-time
tail ($k=l+1$ ) at FNI explicitly. In order to do so, we first calculate
the relevant $\beta$,
\[
\beta_{l+1}\equiv\beta(l,k=l+1)=(-1)^{l+1}\frac{(l+1)(l!)^{2}}{(2l+1)!}=(-1)^{l+1}\frac{2\left[(l+1)!\right]^{2}}{(2l+2)!},
\]
and then obtain {[}using Eq. (\ref{eq:general_late-1}){]}
\begin{equation}
\Psi_{0}^{k=l+1}(u,v)=\hat{c}_{l+1}R^{l+1}(-1)^{l+1}\frac{2\left[(l+1)!\right]^{2}}{(2l+2)!}u^{-(l+1)}\qquad\quad(v\rightarrow\infty,\; u\geq w_{1}).\label{eq:psi_0_FNI}
\end{equation}

\subsubsection{Decay rate at fixed $r_{*}$}

Once the behavior of $\Psi_{0}$ along FNI is known, we can analyze
$\Psi_{0}(u,v)$ throughout the domain $u\geq w_{1}$ and obtain its
decay rate along lines of constant $r_{*}>0$. 

This problem is simplified by the fact that throughout the relevant
domain $v>u\geq w_{1}$, the only relevant piece of the initial null
ray $u=0$ is $v\geq w_{1}$, wherein $\hat{\Psi}_{v}^{k}(v)$ is
an inverse power. Recalling Eqs. (\ref{eq:general_late}) and (\ref{eq:late_explicit}),
it is straightforward to show that $\Psi_{0}(u,v)$ must take the
form $u^{-k}F_{lk}(u/v)$, where $F_{lk}$ is a certain function of
its argument. The problem of determining $\Psi_{0}(u,v)$ then reduces
to that of obtaining the function $F_{lk}(u/v)$. This, in turn, is
achieved by recalling that this function must satisfy a certain ODE,
obtained by substituting the above expression for $\Psi_{0}(u,v)$
in the field equation. Solving this ODE, and matching to a regularity
condition at $r_{*}=0$ (namely $u/v=1$) as well as to the known
boundary condition at FNI (i.e. $u/v=0$), will fully determine $F_{lk}$
and hence also $\Psi_{0}(u,v)$. 

The aforementioned ODE is simplified if we change its independent
variable from $u/v$ to $1-u/v\equiv y$. Thus, we write
\begin{equation}
\Psi_{0}(u,v)=u^{-k}G(y)\qquad\quad(v>u\geq w_{1}),\label{eq:general_late_power}
\end{equation}
where $G$ is a yet unknown function of its argument. The field equation
(\ref{eq:Psi0,uv}) then reduces to the ODE: 
\begin{equation}
G''(y)=\frac{1-k}{1-y}G'(y)+\frac{l(l+1)}{(1-y)y^{2}}G(y).\label{eq:G_ODE}
\end{equation}
This equation is solvable for each $l,k$. However, we shall not need
the explicit solution here, and a general discussion of the properties
of $G(y)$ will suffice (below, however, we shall consider in more
detail the explicit solution in the important case $k=l+1$). 

The limit $y\rightarrow0$ corresponds to $r_{*}=0$. The asymptotic
behavior at this limit can be easily analyzed. The right-hand side
of the above ODE reduces to 
\[
(1-k)G'(y)+\frac{l(l+1)}{y^{2}}G(y).
\]
The term $\propto G'(y)$ becomes unimportant at this limit, and we
obtain the two usual spherical-harmonic asymptotic solutions: a regular
solution $G\simeq y^{l+1}$ and a singular solution $G\simeq y^{-l}$.
\footnote{In both solutions, the additional $\propto G'(y)$ term, and also
the factor $1-y$ in the denominators in Eq. (\ref{eq:G_ODE}), only
affect higher-order terms in the expansion in $y$. %
} The regularity condition at $r_{*}=0$ then selects the regular solution,
yielding 
\begin{equation}
G(y)=g_{0}y^{l+1}+O(y^{l+2})\qquad\quad(y\ll1).\label{eq:G(y)_asymptotic}
\end{equation}
The coefficient $g_{0}$ is to be determined by matching to FNI, but
we shall not need it here. 

Consider now the behavior of $\Psi_{0}$ along a line of constant
$r_{*},$ at very late time ($t\gg r_{*}$). Since $t/r_{*}=2/y-1$,
the inequality $t\gg r_{*}$ corresponds to $y\ll1$; hence, 

\[
\Psi_{0}(u,v)=g_{0}u^{-k}\left[y^{l+1}+O(y^{l+2})\right]\qquad\quad(t\gg r_{*}).
\]
Noting that $y=2r_{*}/(r_{*}+t)\cong2r_{*}/t$, we obtain 
\begin{equation}
\Psi_{0}=\tilde{g}_{0}r_{*}^{\: l+1}\, t^{-(k+l+1)}\left[1+O(r_{*}/t)\right]\qquad\quad(t\gg r_{*}),\label{general-k}
\end{equation}
where $\tilde{g}_{0}\equiv2^{k+l+1}g_{0}$. 

We see that the asymptotic behavior at large $t$ is $\propto t^{-l-k-1}$
and is hence dominated by the smallest possible $k$. However, we
already know that for any $k\leq l$ the field $\Psi_{0}$ vanishes
throughout the domain---namely, at any fixed $r_{*}>0$ it vanishes
for sufficiently large $t$ (stated in other words, the prefactor
$g_{0}$ vanishes for any $k\leq l$). We therefore conclude that
the late-time decay of $\Psi_{0}$ along lines of fixed $r_{*}$ is
dominated (like that at FNI) by the $k=l+1$ term, and decays as $\propto t^{-(2l+2)}$. 

Let us consider in more detail the dominant contribution $ $$k=l+1$.
In this case, the regular solution of Eq. (\ref{eq:G_ODE}) is \emph{exactly}
$G(y)=g_{0}^{(l+1)}y^{l+1}$ (where $g_{0}^{(l+1)}$ is the constant
$g_{0}$ for the case $k=l+1$); hence,
\begin{equation}
\Psi_{0}^{k=l+1}(u,v)=g_{0}^{(l+1)}(y/u)^{l+1}=g_{0}^{(l+1)}\left(\frac{1}{u}-\frac{1}{v}\right)^{l+1}\qquad\quad(v>u\geq w_{1}).\label{eq:Large-t_k0}
\end{equation}
Note the regularity of this expression at $r_{*}=0$. Also, at FNI
it becomes $g_{0}^{(l+1)}u^{-(l+1)}$, which allows us to obtain $g_{0}^{(l+1)}$
by matching to Eq. (\ref{eq:psi_0_FNI}):
\[
g_{0}^{(l+1)}=\hat{c}_{l+1}R^{l+1}\beta_{l+1}=\hat{c}_{l+1}R^{l+1}(-1)^{l+1}\frac{2\left[(l+1)!\right]^{2}}{(2l+2)!}.
\]
To find the asymptotic behavior at $t\gg r_{*}$, recall that 
\[
\frac{1}{u}-\frac{1}{v}=\frac{r_{*}}{uv}\cong\frac{4r_{*}}{t^{2}},
\]
hence
\begin{equation}
\Psi_{0}^{k=l+1}\cong g_{0}^{(l+1)}(4r_{*})^{l+1}\, t^{-(2l+2)}\left[1+O(r_{*}/t)\right]\qquad\quad(t\gg r_{*}).\label{eq:Large-t_Psi0}
\end{equation}
Note that this result {[}Eq. (\ref{eq:Large-t_Psi0}){]} is in complete
correspondence with the one obtained above for general $k$ {[}Eq.
(\ref{general-k}){]}.

\section{Back to the original field $\psi_{l}$ and summary}

So far we have analyzed the late-time tail associated with $\Psi_{(0)}$.
One may wonder what would be the contribution to the tail from the
higher-order moments $\Psi_{(n>0)}$. For example, as noted in the
final part of subsection IVC, one may worry that some $\Psi_{(n>0)}$
would decay at late times slower than $\Psi_{(0)}$ (even though its
amplitude is smaller when $R\gg M$). To this end one may extend the
procedure of \cite{Barack1} (the ``shell toy-model'') from the
Schwarzschild case to our problem. Within that context, it was found
by A. Ori \cite{Amos} that the tail contribution from $\Psi_{(n=1)}$
(in the domain $r_{*}\gg M$) again decays as $\Psi_{(0)}$, though
with a coefficient which scales as $M/R$. There is evidence (emerging
from the horizon's exact conserved quantities discussed below) suggesting
that the mere effect of the $n>0$ term would be to replace the coefficient
$g_{0}^{(l+1)}$ in Eqs. (\ref{eq:Large-t_Psi0},\ref{eq:Large-t_k0})
by another coefficient $\hat{g}$. Note that $\hat{g}\cong g_{0}^{(l+1)}$
as long as $R\gg M$, but in the case of ``strong initial data'',
namely $R\sim M$, one expects that $\hat{g}$ will involve significant
contributions from $n>0$. We shall probably treat this issue further
in a subsequent paper \cite{long} that I hope to address. 

Noting that (i) in the case $R\gg M$ considered here and for evaluation
points in the vicinity of FNI (i.e. at late times and in the domain
$r_{*}\gg M$), $\Psi_{l}$ should be well approximated by $\Psi_{0}$
and that (ii) the late-time decay of $\Psi_{0}$ along lines of fixed
$r_{*}$ is dominated (like that at FNI) by the $k=l+1$ term, we
can write the following expression for the field $\Psi_{l}$ {[}using
Eq. (\ref{eq:Large-t_Psi0}){]}: 
\begin{equation}
\Psi_{l}\cong\hat{g}(4r_{*})^{l+1}\, t^{-(2l+2)}\left[1+O(r_{*}/t)\right]\qquad\quad(t\gg r_{*}\gtrsim R\gg M),\label{eq:Large-t_Psi}
\end{equation}
or {[}using Eq. (\ref{eq:Large-t_k0}){]}:
\begin{equation}
\Psi_{l}\cong\hat{g}\left(\frac{1}{u}-\frac{1}{v}\right)^{l+1}\qquad\quad(v\gg u\gg w_{1}),\label{eq:Psi_nFNI}
\end{equation}
where $\hat{g}\cong g_{0}^{(l+1)}$ (presumably with corrections of
order $M/R$). 

We can now reinvert $\Psi_{l}$ back to the original $\psi_{l}$,
using $\psi_{l}=\hat{T}(\Psi_{l})$, recovering our original problem
of a horizon-based initial perturbation. Recall that in this transformation---which
switches between the horizon and FNI---$t$ is preserved, $r_{*}$
changes its sign, and $u,v$ interchange. The initial-data support
is now restricted to the domain $u>w$ along the ingoing initial ray
($v=0$). This corresponds to $-r_{*}>w$ and in terms of $r$ to
$r\leq M+\delta R$, where $\delta R$ is given by $\delta R=r(r_{*}=-w)-M$
and at the limit $R\gg M$ by $\delta R=M^{2}/R$. The weak-field
condition $R\gg M$ is now mapped into $\delta R\ll M$, to which
we may refer as the ``inverted weak-field'' condition. Therefore,
inverting Eqs. (\ref{eq:Large-t_Psi},\ref{eq:Psi_nFNI}) we get
\begin{equation}
\psi_{l}\cong\hat{g}(-4r_{*})^{l+1}\, t^{-(2l+2)}\left[1+O(r_{*}/t)\right]\qquad\quad(t\gg-r_{*}\gtrsim R\gg M),\label{eq:psi_app_rt}
\end{equation}
and
\begin{equation}
\psi_{l}\cong\hat{g}\left(\frac{1}{v}-\frac{1}{u}\right)^{l+1}=(-1)^{l+1}\hat{g}\left(\frac{1}{u}-\frac{1}{v}\right)^{l+1}\qquad\quad(u\gg v\gg w_{1}).\label{eq:psi_app_uv}
\end{equation}
Note that it is easy to see from Eq. (\ref{eq:psi_app_uv}) that along
the horizon ($u\rightarrow\infty$), $\psi_{l}$ decays as $v^{-l-1}$.

So far, our analysis of $\psi_{l}$ was restricted to evaluation points
in the domain of large $-r_{*}/M$ (and large $r_{*}/M$ for the inverted
field $\Psi_{l}$) and to the case of weak field initial data (for
$\Psi_{l}$. In the case of $\psi_{l}$, the condition is the ``inverted
weak-field'' one), i.e. $R\gg M$ (or $\delta R\ll M$ for $\psi_{l}$).
In the next paper \cite{long}, we would relax this restrictions and
derive an expression for the late-time tail of the field $\psi_{l}$
to the leading order in $|r_{*}|/t$, at any $r_{*}$ and for general
horizon-based initial data (finite $\delta R/M$). In what follows,
\emph{we only sketch very briefly} the key ingredients and main results
and discuss the case of general initial data and relations to previous
works.

We begin with the ``inverted weak-field'' restriction $\delta R\ll M$.
We would like to relax this assumption and to achieve better control
on the contribution of higher-order terms in the iteration scheme
(which presumably yields contributions $\propto(\delta R/M)^{n}$
to $\hat{g}$). As was mentioned above, an analysis of the $n=1$
term was already carried out (in the $\Psi_{l}$ context), indicating
that the effect of this term is to merely modify the coefficient $g_{0}^{(l+1)}$,
by an amount $\propto M/R$. We would like in addition (i) to verify
that this behavior indeed extends to higher-$n$ terms as well and
(ii) to obtain the explicit expression for $\hat{g}$, for \emph{finite}
$\delta R/M$, which incorporates the overall contributions of all
$n$. 

To achieve these goals we exploit the conserved quantities found by
Aretakis \cite{Aretakis}. He analyzed the evolution of certain derivatives
of $\psi_{l}$ along the horizon of extremal RN, and obtained interesting
exact results, which are easiest to express in ingoing-Eddington ($v,r$)
coordinates. Following are the results which are most relevant to
our analysis: (i) The quantity $A_{l}=d_{l}(\psi_{l})$, where $d_{l}$
is the differential operator $(\partial/\partial r)^{l+1}+(l/M)(\partial/\partial r)^{l}$
evaluated at the horizon, is \emph{exactly conserved} (i.e. independent
of $v$), and (ii) for all $j\leq l$, $(\partial/\partial r)^{j}\psi_{l}$
decays to zero as $v\rightarrow\infty$. (Note that (ii) also implies
that $(\partial/\partial r)^{l+1}\psi_{l}$ asymptotically approaches
$A_{l}$ at $v\rightarrow\infty$.)

By applying the operator $d_{l}$ to the \emph{original} initial data
(\ref{eq:Horizon_Expansion}) (i.e. before the inversion and the insertion
of the scale parameter $R$) and to Eq. (\ref{eq:psi_app_uv}), one
verifies the consistency of the latter asymptotic expression, and
furthermore obtains the exact coefficient $\hat{g}$:
\begin{equation}
\hat{g}=M^{l+1}(-1)^{l+1}\frac{2\left[(l+1)!\right]^{2}}{(2l+2)!}\left(c_{l+1}+\frac{l}{l+1}c_{l}\right)=M^{l+1}\beta_{l+1}\left(c_{l+1}+\frac{l}{l+1}c_{l}\right).\label{eq:g_hat}
\end{equation}
Note that this is an \emph{exact} expression, not restricted to the
``inverted weak-field'' case.

We now turn to discuss the second restriction, according to which
our calculation of $\psi_{l}$ is a good approximation only for evaluation
points in the domain of large $-r_{*}/M$. One might suspect that
once $-r_{*}$ increase and becomes comparable to (or smaller than)
$M$---and obviously when $r_{*}$ becomes positive---this approximation
would break down. In the subsequent paper \cite{long} we shall show,
however, that in fact the validity of the late-time behavior $t^{-(2l+2)}$
is not restricted to the domain $M\ll-r_{*}$. Rather, it holds throughout
the domain $t\gg|r_{*}|$. The only thing which changes when $-r_{*}$
becomes comparable to (or smaller than) $M$ is the dependence on
$r_{*}$: The power law $(-r_{*})^{l+1}$ is replaced by a static
solution of the field equation. 

In order to obtain this result, we employ the so-called \emph{late-time
expansion}, presented for example in Refs. \cite{Late-time,Barack2}.
In this procedure, we assume that $\psi_{l}(r,t)$ admits a large-$t$
asymptotic behavior of the form $\sum_{j=0}^{\infty}H_{j}(r)(t/M)^{-m-j}$,
with a certain leading power $m$ and certain functions $H_{j}(r)$.
Substituting this expression in the field equation, one first finds
that $H_{j=0}(r)$ must be a static solution (regular at $r\rightarrow\infty$),
namely $\alpha(r/M)(r/M-1)^{-l-1}$, where $\alpha$ is a constant.
In addition, one obtains a hierarchy of ODEs for the various functions
$H_{j>0}(r_{*})$ (whose forms will not concern us here). To obtain
the leading power $m$ (along with $\alpha$), one matches the near-horizon
asymptotic form of the dominant term $j=0$ ---namely $\alpha(-r_{*}/M)^{l+1}(t/M)^{-m}$
--- to the large-$t$ asymptotic form of $\psi_{l}$ obtained in Eq.
(\ref{eq:psi_app_rt}). The matching zone is $t\gg-r_{*}\gg M$. One
then easily finds that $m=2l+2$ and $\alpha=4^{l+1}M^{-(l+1)}\hat{g}$.
Therefore, the final result is 
\begin{equation}
\psi_{l}(r,t)\cong4^{l+1}\hat{g}M^{l+1}(r/M)(r/M-1)^{-l-1}\, t^{-2l-2}\qquad\quad(t\gg|r_{*}|).\label{eq:Final}
\end{equation}
Note that $\hat{g}$ contains a factor of $M^{l+1}$ and is therefore
dimensionful {[}see Eq. (\ref{eq:g_hat}){]}. 

So far we have addressed the case of horizon-based initial data. In
the other case of off-horizon compact initial data (case I in the
classification above), all coefficients $c_{k}$ vanish, and so does
$\hat{g}$. The decay rate will then be faster than $t^{-2l-2}$.
A prototype of that case is the situation of a (compact-support) pulse
incident from large $r$ towards the BH. In this case the classic
analysis by Price \cite{Price} should apply. This analysis (like
subsequent variants \cite{Barack1,Barack2}) demonstrates that the
Schwarzschild's large-$t$ tail is predominantly a weak-field phenomenon---that
is, a scattering of a certain ingoing field off a weak-field curvature
potential. The latter is $\propto M$ and is insensitive to $Q$ (which
only affects the curvature at the next-to-leading order). Thus, in
the case of off-horizon compact initial data we expect a large-$t$
tail $\propto t^{-2l-3}$---and an FNI tail $\propto u^{-l-2}$---just
like in Schwarzschild. Note, however, that since ingoing perturbations
will usually cross the horizon, due to the presence of centrifugal-like
potential there, a scattering dynamics will take place near the horizon
as well (similar to that at large $r$), and will lead to an inverse-power
decay along the horizon, as $v^{-l-2}$. This also follows quite immediately
from the fact that the notion of ``off-horizon compact initial data''
is invariant under $T$-inversion, and the latter maps the standard
$\propto u^{-l-2}$ decay at FNI to a $\propto v^{-l-2}$ decay at
the horizon. 

As was mentioned above, a generic compact-support initial data may
be represented as a superposition of the two types, namely ``off-horizon''
and ``horizon-based'' initial-data sets. The late-time decay will
then be dominated by the above ``horizon-based'' contribution: Decay
rates like $u^{-l-1}$ at FNI, like $v^{-l-1}$ along the horizon,
and like $t^{-2l-2}$ at large $t$ (namely $t\gg|r_{*}|$), according
to Eq. (\ref{eq:Final}), along with (\ref{eq:g_hat},\ref{eq:psi_app_uv}). 

In Ref. \cite{Bicak}, Bicak obtained a large-$t$ asymptotic behavior
which involves the same function of $r$ as in (\ref{eq:Final}),
but with a different inverse power $t^{-l-2}$. For $l\neq0$, his
result is inconsistent with our analysis. Furthermore, one can use
Bicak's expression as a seed of a large-$t$ expansion, constructing
the corresponding $\psi_{l}$ (with $m=l+2$ rather than $2l+2$)
\cite{long}, and when $d_{l}$ is applied to this $\psi_{l}$, one
finds that $A_{l}$ actually diverges, which is inconsistent with
regular initial data. The numerical results \cite{Lucietti} for $l=1,2$
also excludes the inverse power $t^{-l-2}$ and support our analytical
result $t^{-2l-2}$. It remains unclear where for $l\neq0$ the problem
in \cite{Bicak} arose.

More generally, by combining the large-$t$ expansion and the conservation
of $A_{l}$, one can show that for generic regular initial data, the
only consistent inverse power is $m=2l+2$ \cite{long}: $A_{l}$
diverges for any $m\leq2l+1$, and vanishes for any $m\geq2l+3$ (which
would be inconsistent with generic regular initial data, wherein $\hat{g}\neq0$). 

In Ref. \cite{Burko}, Blaksley and Burko investigated two special
cases of initial data: (i) initial static moment that extends up to
the horizon, and (ii) the case of ``off-horizon compact initial support''
(in our terminology). In both cases, all the coefficients $c_{k<l}$
in the initial-data horizon expansion (\ref{eq:Horizon_Expansion})
vanish. Obviously, analyzing such a subclass of initial data is inadequate
for testing Bicak's claim, especially because a-priori one might naturally
expect that the smallest-$k$ terms would dominate the late-time tail.
When restricted to that subclass, our results are consistent with
those of Blaksley and Burko. In particular, for an initial static
moment at the horizon, one finds that $c_{l+1},c_{l}\neq0$ (and all
other $c_{k}$ vanish), yielding $\hat{g}\neq0$ and hence a late-time
decay $\propto t^{-2l-2}$. 

It may be interesting to extend this analysis to coupled gravitational
and electromagnetic perturbations on extremal RN. In Ref. \cite{Bicak2},
Bicak employed his results from \cite{Bicak3} and showed that the
scalar-field perturbations serve as a prototype for these coupled
perturbations; he then used his results from Ref. \cite{Bicak} to
deduce the late-time behavior of them. It will therefore be interesting
to revisit his analysis in \cite{Bicak2} and obtain the late-time
behavior of coupled perturbations (on ERN) using the results of the
present paper.

\section*{ACKNOWLEDGMENTS}

I would like to thank Professor A. Ori for his guidance throughout
the execution of this research and for countless helpful discussions.

\end{document}